\documentclass[sigconf]{acmart}

\usepackage{balance}
\usepackage{amsmath,amsfonts,amsthm}
\usepackage{algorithm}
\usepackage{algpseudocode}
\usepackage{array}
\usepackage{textcomp}
\usepackage{stfloats}
\usepackage{url}
\usepackage{verbatim}
\usepackage{graphicx}
\usepackage{threeparttable}
\usepackage{bibentry}
\usepackage{multirow}
\usepackage{epsfig}
\usepackage{epstopdf}
\usepackage{ragged2e}
\usepackage{rotating}
\usepackage{float}
\usepackage{multicol}
\usepackage{bbding}
\usepackage{microtype}
\usepackage{booktabs}
\usepackage{xcolor}
\usepackage[many]{tcolorbox}
\usepackage{mathtools}
\usepackage{etoolbox}
\usepackage{cases}
\usepackage{enumitem}
\usepackage{xspace}
\newcolumntype{C}[1]{>{\centering\arraybackslash}p{#1}}

\makeatletter
\DeclareRobustCommand\onedot{\futurelet\@let@token\@onedot}
\def\@onedot{\ifx\@let@token.\else.\null\fi\xspace}

\makeatother

\AtBeginDocument{%
  }

\setcopyright{acmlicensed}
\copyrightyear{2027}
\acmYear{2027}
\setcopyright{cc}
\setcctype{by}
\acmConference[WSDM '27]{Proceedings of the Twentieth ACM International Conference on Web Search and Data Mining}{February 15--19, 2027}{Hong Kong, Hong Kong}
\acmBooktitle{Proceedings of the Twentieth ACM International Conference on Web Search and Data Mining (WSDM '27), February 15--19, 2027, Hong Kong, Hong Kong}
\acmDOI{10.1145/3862322.3863719}
\acmISBN{979-8-4007-3117-4/2027/02}

\begin{document}

\title{RAGR: Review-Augmented Generative Recommendation}

\author{Yingyi Zhang}
\orcid{0000-0001-9062-3428}
\affiliation{
  \institution{Dalian University of Technology \& City University of Hong Kong}
  \city{Dalian}
  \country{China}
}
\email{yingyizhang@mail.dlut.edu.cn}

\author{Junyi Li}
\orcid{0009-0007-0480-5593}
\author{Yejing Wang}
\orcid{0000-0003-2852-9910}
\affiliation{
  \institution{City University of Hong Kong}
  \city{Hong Kong}
  \country{China}
}
\email{junyili@cityu.edu.hk}
\email{yejing.wang@my.cityu.edu.hk}

\author{Wenlin Zhang}
\orcid{0000-0003-1809-8264}
\author{Xiaowei Qian}
\orcid{0009-0007-6448-0890}
\affiliation{
  \institution{City University of Hong Kong}
  \city{Hong Kong}
  \country{China}
}
\email{wl.z@my.cityu.edu.hk}
\email{xiaowqian2-c@my.cityu.edu.hk}

\author{Sheng Zhang}
\orcid{0009-0006-1758-6708}
\author{Binhao Wang}
\orcid{0009-0009-6671-3613}
\affiliation{
  \institution{City University of Hong Kong}
  \city{Hong Kong}
  \country{China}
}
\email{szhang844-c@my.cityu.edu.hk}
\email{binhao.wang@my.cityu.edu.hk}

\author{Yue Feng}
\orcid{0000-0001-7741-8534}
\affiliation{
  \institution{Dalian University of Technology}
  \city{Dalian}
  \country{China}
}
\email{fy\_0403@mail.dlut.edu.cn}

\author{Yichao Wang}
\orcid{0000-0001-7053-8269}
\author{Yong Liu}
\orcid{0000-0001-9031-9696}
\affiliation{
  \institution{Huawei Technologies Ltd.}
  \city{Singapore}
  \country{Singapore}
}
\email{wangyichao5@huawei.com}
\email{liu.yong6@huawei.com}

\author{Kaidi Xu}
\orcid{0000-0003-4437-0671}
\affiliation{
  \institution{City University of Hong Kong}
  \city{Hong Kong}
  \country{China}
}
\email{kaidixu@cityu.edu.hk}

\author{Xiangyu Zhao}
\orcid{0000-0003-2926-4416}
\correspondingauthor
\affiliation{
  \institution{City University of Hong Kong}
  \city{Hong Kong}
  \country{China}
}
\email{xianzhao@cityu.edu.hk}

\author{Xianneng Li}
\orcid{0000-0003-4130-6930}
\correspondingauthor
\affiliation{
  \institution{Dalian University of Technology}
  \city{Dalian}
  \country{China}
}
\email{xianneng@dlut.edu.cn}

\renewcommand{\shortauthors}{Zhang et al.}

\begin{abstract}

Sequential recommendation (SR) is traditionally formulated as next-item prediction over chronological interactions. Although recent generative recommendation (GR) methods introduce semantic IDs, autoregressive decoding, and unified token spaces, they largely inherit this item-only assumption. We argue that this design constitutes a structural bottleneck, because user decision-making is not purely behavioral: while item interactions reveal what users choose, review feedback often explains why they choose it by exposing latent evaluative factors.

Motivated by this, we propose \textbf{R}eview-\textbf{A}ugmented \textbf{G}enerative \textbf{R}ecommendation (\textbf{RAGR}), a framework integrating reviews directly into the generative user sequence rather than treating them as auxiliary side information. Specifically, \emph{Review-Augmented User Sequence Modeling} chronologically interleaves item and review semantic IDs into a mixed behavioral-semantic sequence, enabling review token to participate directly in autoregressive next-token generation. To preserve the recommendation objective, an \emph{Item-Centric Task Generation Alignment} strategy based on direct preference optimization (DPO) encourages the model to favor item tokens over review tokens at prediction positions. Experiments on three real-world datasets demonstrate consistent, significant gains over strong GR backbones. Code is available at \url{https://github.com/Applied-Machine-Learning-Lab/WSDM2027_RAGR}.

\end{abstract}



\maketitle

\section{Introduction}

Recommender systems are indispensable to modern e-commerce platforms~\cite{zhao2024recommender,wang2026embedding}, such as Amazon~\cite{he2016ups} and Alibaba~\cite{wang2018billion}, where they surface relevant items from catalogs of millions.
Among various paradigms, sequential recommendation (\textbf{SR})~\cite{kang2018self,xie2022contrastive,wang2026multi, liu2025sigma} stands out because user preferences are inherently \emph{dynamic}, shaped not only by static profiles but by the evolving temporal patterns of historical interactions~\cite{zhou2018deep,zhou2019deep, zhao2018deep,zhao2018recommendations,zhang2023meta}.
SR is predominantly cast as a \emph{next-item prediction} task---given a chronological sequence of past interactions, the model forecasts the item a user will engage with next~\cite{pan2026survey,fang2020deep}. 
This formulation is tightly coupled with the representation mechanism of conventional models: architectures such as SASRec~\cite{kang2018self} and BERT4Rec~\cite{sun2019bert4rec} rely on an \emph{item ID embedding table}, under which a user's history is encoded as an ID sequence and the learning objective reduces to predicting the next item ID.

\begin{figure}[t!]
\centering
{\includegraphics[width=0.99\linewidth]{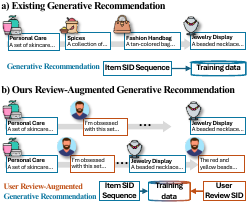}}
\caption{Comparison Between Existing and Review-Augmented Generative Recommendation.}
\label{fig:intro}
\vspace{-15pt}
\end{figure}

With the emergence of generative recommendation (\textbf{GR})~\cite{tiger,li2025survey}, this paradigm has been reparameterized through semantic ID (SID)~\cite{tiger,letter,tokenrec}, autoregressive decoding, and unified token spaces. GR replaces the conventional item-ID lookup table with semantic tokenization and generation. 
However, despite this representational shift, the underlying modeling paradigm remains essentially the same as that in Fig.~\ref{fig:intro} a): existing GR methods still model the user history as an item-only trajectory and optimize for generating the next item.
We argue that this inherited assumption is not merely a missed opportunity but a \emph{structural bottleneck}.  
In reality, user decision-making is inherently multi-faceted: purchases are often preceded by browsing and comparison, and followed by explicit feedback such as textual reviews~\cite{zheng2017joint, hai2013identifying}. 
As McAuley and Leskovec~\cite{mcauley2013hidden} show, reviews expose \emph{hidden evaluative dimensions}---such as quality, usability, and aesthetics---that shape a user's rating but remain invisible in interaction logs alone. An item-only sequence captures only the \emph{outcome} of a decision---\emph{what} the user chose---while omitting the explanatory signals that reveal \emph{why} it was chosen~\cite{zhao2023kuaisim,zhao2021usersim, zhang2023denoising,fan2023adversarial}. 
Without such signals, the model must infer preference dynamics solely from item co-occurrence patterns, resulting in a shallow and brittle understanding of user intent.

Notably, the GR framework possesses the machinery needed to transcend this bottleneck.
A distinguishing property of GR is its \emph{unified tokenizer}, which projects item text into a shared semantic space~\cite{li2025survey, deldjoo2024review, wang2025rethinking}, allowing recommendation objects to be represented and generated jointly---a capability conventional ID-based recommendation lacks.
In current practice, however, this tokenizer is applied almost exclusively to items~\cite{tiger,letter,tokenrec}, while the same mechanism could equally operate on other textual signals users routinely produce---most notably post-interaction reviews, which carry fine-grained preference signals beyond item transitions.
In short, prior work has advanced the expressiveness of \emph{item representation} yet left the \emph{user sequence} largely untouched.
This asymmetry gives rise to our central research question:
\textbf{\textit{Can the unified tokenizer of GR be repurposed to encode review feedback as semantic tokens and weave them into the interaction sequence, enabling the model to capture not only behavioral outcomes but also the underlying preference rationale?}}
Addressing this entails two intertwined challenges:
\textit{(1)~Heterogeneous sequence construction}---items and reviews differ in granularity and functional role, so projecting both through a shared tokenizer into a coherent sequence risks disrupting the sequence's positional semantics and injecting noise.
\textit{(2)~Recommendation-objective preservation}---once review tokens enter the sequence, a principled alignment mechanism must ensure they serve as supporting evidence for next-item prediction rather than a competing generation target.

To address these challenges, we propose \emph{\textbf{R}}eview-\emph{\textbf{A}}ugmented \emph{\textbf{G}}enerative \emph{\textbf{R}}ecommendation (\textbf{RAGR}), as illustrated in Fig.~\ref{fig:intro} b).  
The central idea of RAGR is to move beyond the conventional treatment of reviews as auxiliary features in discriminative recommenders. Instead, we bring review feedback directly into the GR paradigm by encoding reviews as tokens within the user sequence itself—extending the item-only behavioral trajectory into a review-augmented one.
Specifically, RAGR consists of two complementary components.
\textbf{(i)~Review-Augmented User Sequence Modeling.} Item SIDs and review SIDs are interleaved chronologically to form a mixed-signal user sequence, over which the GR backbone is trained to predict the next SID---allowing review context to directly participate in the autoregressive generation process.
\textbf{(ii)~Item-Centric Task Generation Alignment.} Because the mixed sequence may cause the model to devote generation capacity to reviews rather than items, we apply a direct preference optimization (DPO)~\cite{dpo} alignment mechanism that teaches the model to \emph{prefer} generating item SIDs over review SIDs at prediction positions, ensuring that the learning objective remains anchored to next-item recommendation.

Our main contributions are summarized as follows:
\begin{itemize}[leftmargin=*]
    \item We identify a structural limitation of existing GR methods—their confinement to item-only sequences—and propose \textbf{RAGR}, to our knowledge the first framework to elevate reviews from auxiliary features to tokens within the GR paradigm, extending user sequences from purely behavioral to mixed behavioral-semantic.
    \item We design a \emph{Review-Augmented User Sequence Modeling} mechanism that chronologically interleaves item and review SIDs, letting review context enrich the user sequence.
    \item We propose an \emph{Item-Centric Task Generation Alignment} strategy based on DPO, ensuring reviews inform---but never displace---the next-item objective.
    \item We conduct extensive experiments on three real-world datasets, showing that RAGR yields consistent gains across  GR backbones.
\end{itemize}

\section{Problem Formulation}
\label{sec:defination}

Let $\mathcal{U}$ and $\mathcal{I}$ denote the user set and item set, respectively. For each user $u \in \mathcal{U}$, we observe a chronological interaction history $\mathcal{S}_u = \big[(i_1, r_1), (i_2, r_2), \dots, (i_T, r_T)\big],$
where $i_t \in \mathcal{I}$ denotes the interacted item at step $t$, and $r_t$ denotes the associated textual feedback, such as a review. Different from conventional sequential recommendation, which models only the item sequence $(i_1,i_2,\dots,i_T)$, we consider a review-augmented user sequence in which item interactions and review feedback are jointly modeled. Given the historical sequence up to step $t-1$,
\begin{equation}
\mathcal{S}_u^{<t} = \big[(i_1, r_1), (i_2, r_2), \dots, (i_{t-1}, r_{t-1})\big],
\end{equation}
our goal is to predict the next target item $i_t$. Formally, we aim to learn a generative recommendation model
$f_\theta : \mathcal{S}_u^{<t} \mapsto i_t$, where $\theta$ denotes model parameters.

To support generative modeling, we represent both items and reviews in a unified token space. Let $\mathbf{z}(i_t)$ and $\mathbf{z}(r_t)$ denote the tokenized representations of item $i_t$ and review $r_t$, respectively. Then the historical sequence can be rewritten as
\begin{equation}
\widetilde{\mathcal{S}}_u^{<t}
=
\big[\mathbf{z}(i_1), \mathbf{z}(r_1), \mathbf{z}(i_2), \mathbf{z}(r_2), \dots, \mathbf{z}(i_{t-1}), \mathbf{z}(r_{t-1})\big].
\end{equation}
Accordingly, the recommendation objective is to generate the tokenized representation of the next target item: $p_\theta\big(\mathbf{z}(i_t)\mid \widetilde{\mathcal{S}}_u^{<t}\big)$.

Based on this formulation, we construct the training set as
\begin{equation}
\mathcal{D}
=
\left\{
\big(\widetilde{\mathcal{S}}_u^{<t}, \mathbf{z}(i_t), \mathbf{z}(r_t)\big)
\,\middle|\,
u \in \mathcal{U},\; 2 \le t \le T
\right\},
\end{equation}
where $\mathbf{z}(i_t)$ is the tokenized target item and $\mathbf{z}(r_t)$ is the corresponding review of the user.
The learning problem in this work is therefore to perform next-item generation over review-augmented user sequences while preserving the item-centric recommendation objective. \textbf{\textit{However, incorporating review tokens into the unified generative space introduces ambiguity in the prediction target, since both item sequences and review sequences become plausible continuations under the same context.} }

To resolve this ambiguity, the proposed RAGR framework
addresses two sub-problems in sequence: (1)~how to train a generative model over the mixed item--review SID sequence so
that review context enriches preference modeling (Section~\ref{sec:review_augmented}), and (2)~how to align the generation objective so that the model favors item SIDs over review SIDs at prediction positions (Section~\ref{sec:task_alignment}).

\section{The proposed RAGR method}
\label{sec:methodlogy}

\begin{figure*}[ht]
\centering
\includegraphics[width=\textwidth]{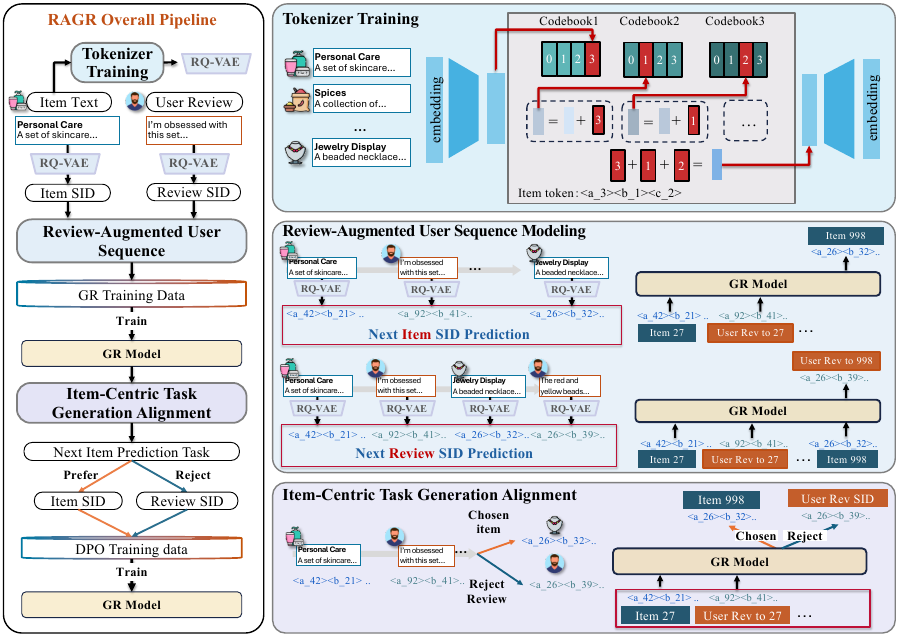}
\caption{The overall framework of the proposed \textbf{RAGR}, which consists of three main stages: Tokenizer Training, Review-Augmented User Sequence Modeling, and Item-Centric Task Generation Alignment.} 
\label{fig:method}
\end{figure*}

\subsection{Overview of RAGR}
As illustrated in Fig.~\ref{fig:method}, RAGR consists of three stages. First, \emph{Tokenizer Training} trains an RQ-VAE~\cite{tiger} tokenizer that maps both items and reviews into a unified SID space, producing the discrete tokens. Second, \emph{Review-Augmented User Sequence Modeling} interleaves item interactions with their associated review into sequences, over which the GR is trained for next-SID generation, conditioning recommendations on both item interactions and semantic evidence from reviews. Third, \emph{Item-Centric Task Generation Alignment} employs DPO-based preference alignment to bias generation toward target items rather than reviews, so that reviews serve as evidence for item selection rather than a competing generation target.

\subsection{Tokenizer Training}
\label{sec:token_train}

To address the need for a unified discrete token space for both item interactions and review feedback, we first train a semantic tokenizer based on item text. As shown in Fig.~\ref{fig:method}, we first encode item text into dense semantic representations with an LLM-based text encoder, and then train an RQ-VAE~\cite{tiger} tokenizer to quantize such representations into multi-level discrete codes. In this way, each item can be represented as a semantic ID sequence that serves as the basis for subsequent review-augmented GR.
Formally, let $x_i$ denote the textual content associated with item $i \in \mathcal{I}$, such as its title, category, or description. We first employ a pretrained LLM-based text encoder to obtain a dense embedding $\mathbf{e}_i = E(x_i)$,
where $E(\cdot)$ denotes the text encoder and $\mathbf{e}_i \in \mathbb{R}^d$ is the semantic embedding of item $i$. Compared with raw item IDs, such text-informed representations provide richer semantic information and serve as the basis for tokenizer learning.

Based on these item embeddings, we train a residual quantization variational autoencoder (RQ-VAE) to discretize the continuous semantic space into a sequence of code indices. Specifically, given an item embedding $\mathbf{e}_i$, the encoder of RQ-VAE first maps it into a latent representation $\mathbf{h}_i = g_{\text{enc}}(\mathbf{e}_i)$,
where $g_{\text{enc}}(\cdot)$ denotes the encoder network. We then apply residual quantization over $M$ codebooks $\{\mathcal{C}^{(1)}, \mathcal{C}^{(2)}, \dots, \mathcal{C}^{(M)}\}$ in a coarse-to-fine manner. At the $m$-th level, the residual representation is quantized by selecting the nearest codeword from the corresponding codebook:
\begin{equation}
q_i^{(m)} = \arg\min_{k} \left\| \mathbf{r}_i^{(m-1)} - \mathbf{c}^{(m)}_k \right\|_2^2,
\end{equation}
where $\mathbf{c}^{(m)}_k \in \mathcal{C}^{(m)}$ denotes the $k$-th codeword in the $m$-th codebook, and $\mathbf{r}_i^{(m-1)}$ is the residual to be quantized at level $m$. The residual is updated as $\mathbf{r}_i^{(m)} = \mathbf{r}_i^{(m-1)} - \mathbf{c}^{(m)}_{q_i^{(m)}}$,
with $\mathbf{r}_i^{(0)} = \mathbf{h}_i$. After $M$ levels of quantization, the item representation is approximated by the sum of selected codewords: $\hat{\mathbf{h}}_i = \sum_{m=1}^{M} \mathbf{c}^{(m)}_{q_i^{(m)}}$.
Accordingly, each item $i$ is represented as a multi-level discrete SID
\begin{equation}
\mathbf{z}(i) = \big[q_i^{(1)}, q_i^{(2)}, \dots, q_i^{(M)}\big],
\end{equation}
which serves as its tokenized representation in the unified generative space. Intuitively, earlier codebooks capture coarse-grained semantics, while later codebooks progressively refine the representation with finer-grained details. Following standard residual quantization objectives~\cite{lee2022autoregressive}, we train the RQ-VAE by minimizing the discrepancy between the latent representation and its quantized approximation, together with codebook and commitment regularization:
\begin{equation}
\resizebox{\linewidth}{!}{$\displaystyle
\mathcal{L}_{\text{tok}}
=\ \underbrace{\left\| \mathbf{h}_i - \hat{\mathbf{h}}_i \right\|_2^2}_{\mathcal{L}_{\text{rec}}} 
+ \underbrace{\sum_{m=1}^{M}\left\| \mathrm{sg}\!\left[\mathbf{r}_i^{(m-1)}\right] - \mathbf{c}^{(m)}_{q_i^{(m)}} \right\|_2^2}_{\mathcal{L}_{\text{code}}} 
+ \gamma \underbrace{\sum_{m=1}^{M}\left\| \mathbf{r}_i^{(m-1)} - \mathrm{sg}\!\left[\mathbf{c}^{(m)}_{q_i^{(m)}}\right] \right\|_2^2}_{\mathcal{L}_{\text{commit}}}
$}.
\end{equation}
where $\mathrm{sg}[\cdot]$ denotes the stop-gradient operator and $\gamma$ is the commitment coefficient.

\subsection{Review-Augmented User Sequence Modeling}
\label{sec:review_augmented}

To address the limitation that existing generative recommendation models primarily capture what users selected but not why they selected it, we augment the conventional item-only interaction sequence with review feedback, so that each historical interaction is represented by both its item semantic ID and its corresponding review semantic ID. As illustrated in Fig.~\ref{fig:method}, instead of organizing user behavior as a pure item trajectory, we interleave item and review tokens in chronological order, allowing the generative recommender to condition sequence modeling on both behavioral outcomes and semantic feedback evidence.
Let $\mathbf{z}(i_t)$ and $\mathbf{z}(r_t)$ denote the semantic IDs of item $i_t$ and review $r_t$, respectively, obtained from the tokenizer trained in the first stage (Section~\ref{sec:token_train}). For each user $u$, we construct the review-augmented sequence as
\begin{equation}
\widetilde{\mathcal{S}}_u
=
\big[
\mathbf{z}(i_1), \mathbf{z}(r_1), \mathbf{z}(i_2), \mathbf{z}(r_2), \dots, \mathbf{z}(i_T), \mathbf{z}(r_T)
\big].
\end{equation}
Compared with conventional item-only sequences, this formulation allows the model to observe not only what the user interacted with, but also how the user described or evaluated those interactions.

Based on the review-augmented sequence, we construct a unified textual generation task by serializing item SIDs and review SIDs into autoregressive training instances. Specifically, we generate two types of sequence-to-sequence samples from the same review-augmented interaction history.
The first type is \emph{next-item SID prediction}, where the input consists of the historical item-review sequence up to step $t-1$, and the target is the semantic ID sequence of the next item $\mathbf{z}(i_t)$. The second type is \emph{next-review SID prediction}, where the input additionally includes the target item $\mathbf{z}(i_t)$, and the target is the corresponding review semantic ID sequence $\mathbf{z}(r_t)$. In both cases, the model is trained as a standard generative recommender over textualized semantic IDs, rather than as two separate task-specific predictors.

Formally, let $(\mathbf{x}, \mathbf{y})$ denote a generic training instance constructed from the review-augmented sequence, where $\mathbf{x}$ is the serialized input sequence and $\mathbf{y}$ is the target semantic ID sequence to be generated. Then the GR model is optimized with a unified autoregressive generation objective:
\begin{equation}
\mathcal{L}_{\text{seq}}
=
-\sum_{(\mathbf{x},\mathbf{y}) \in \mathcal{D}_{\text{seq}}}
\log p_{\theta}(\mathbf{y}\mid \mathbf{x}),
\end{equation}
where $\mathcal{D}_{\text{seq}}$ denotes the collection of all serialized next-item and next-review training instances. Under this formulation, item prediction and review prediction are not treated as separate optimization objectives, but as two forms of textual sequence generation derived from the same review-augmented user behavior.

This unified sequence modeling strategy provides two benefits. First, it exposes the model to a finer-grained generative sequence that interleaves item interactions and review feedback, thereby enriching user sequence modeling beyond pure item transitions. Second, it allows the GR model to learn the semantic dependency between items and their associated feedback within a shared generative space, which lays the foundation for the subsequent item-centric task alignment stage.

\subsection{Item-Centric Task Generation Alignment}
\label{sec:task_alignment}
To address the risk that review-augmented sequence modeling may blur the task boundary of next-item recommendation and inadvertently drive the model to generate review semantic IDs, we further introduce an item-centric task alignment stage that explicitly aligns the GR model toward target item generation rather than review generation. As illustrated in Fig.~\ref{fig:method}, the key idea is to treat the target item semantic ID as the \emph{preferred} output and the corresponding review semantic ID as the \emph{rejected} output under the same context, encouraging the model to assign a higher conditional likelihood to the item SID than the review SID. In this way, review remains informative evidence for item prediction instead of becoming a competing generation target, and this alignment complements the review-augmented sequence modeling stage by keeping the overall objective anchored to next-item prediction.

Formally, let $\mathbf{x}_t$ denote the serialized review-augmented context constructed from the history before the target interaction, i.e.,
\begin{equation}
\mathbf{x}_t = \big[\mathbf{z}(i_1), \mathbf{z}(r_1), \dots, \mathbf{z}(i_{t-1}), \mathbf{z}(r_{t-1})\big].
\end{equation}
For each target interaction at step $t$, we form a preference pair $\big(\mathbf{x}_t,\; \mathbf{y}_t^{+},\; \mathbf{y}_t^{-}\big)$, where
\begin{equation}
\mathbf{y}_t^{+} = \mathbf{z}(i_t), \qquad \mathbf{y}_t^{-} = \mathbf{z}(r_t).
\end{equation}
Here, $\mathbf{y}_t^{+}$ is the preferred output because it corresponds to the next target item that should be recommended, while $\mathbf{y}_t^{-}$ is the rejected output because it corresponds to review feedback, which should inform recommendation but should not replace the recommendation target itself.
Based on these pairs, we construct a training set
\begin{equation}
\mathcal{D}_{\text{dpo}}
=
\left\{
\big(\mathbf{x}_t,\mathbf{y}_t^{+},\mathbf{y}_t^{-}\big)
\,\middle|\,
u \in \mathcal{U},\; 2 \le t \le T
\right\}.
\end{equation}
We then apply DPO~\cite{dpo} to align the GR model toward the preferred item output. Let $\pi_{\theta}$ denote the current GR model and $\pi_{\mathrm{ref}}$ denote a reference model. To improve readability, we first define the relative preference score between the target item and the review signal under the same context as
\begin{equation}
\Delta_{\theta}(\mathbf{x},\mathbf{y}^{+},\mathbf{y}^{-})
=
\log \frac{\pi_{\theta}(\mathbf{y}^{+}\mid \mathbf{x})}
{\pi_{\mathrm{ref}}(\mathbf{y}^{+}\mid \mathbf{x})}
-
\log \frac{\pi_{\theta}(\mathbf{y}^{-}\mid \mathbf{x})}
{\pi_{\mathrm{ref}}(\mathbf{y}^{-}\mid \mathbf{x})}.
\end{equation}
For each preference triple $(\mathbf{x},\mathbf{y}^{+},\mathbf{y}^{-}) \in \mathcal{D}_{\text{dpo}}$, the DPO loss is then written as $\ell_{\text{DPO}}(\mathbf{x},\mathbf{y}^{+},\mathbf{y}^{-})= -\log \sigma\!\big(\beta \, \Delta_{\theta}(\mathbf{x},\mathbf{y}^{+},\mathbf{y}^{-})\big),$
and the overall objective is
\begin{equation}
\mathcal{L}_{\text{DPO}} = \mathbb{E}_{(\mathbf{x},\mathbf{y}^{+},\mathbf{y}^{-}) \sim \mathcal{D}_{\text{dpo}}} \left[ \ell_{\text{DPO}}(\mathbf{x},\mathbf{y}^{+},\mathbf{y}^{-}) \right].
\label{eq:dpo}
\end{equation}
Here, $\sigma(\cdot)$ denotes the sigmoid function, and $\beta$ is a temperature coefficient that controls the sharpness of optimization.

\section{Experiment}
\label{sec:Experiments}

In this section, we answer the following four research questions:

\begin{itemize}[leftmargin=*]
    \item \textbf{RQ1:} To what extent can RAGR improve the performance of existing generative recommendation backbones on next-item recommendation?
    \item \textbf{RQ2:} How effective are the two key components of RAGR, namely \emph{Review-Augmented User Sequence Modeling} and \emph{Item-Centric Task Generation Alignment}?
    \item \textbf{RQ3:} Which tokenizer training strategy is most effective for RAGR: training RQ-VAE on item text only, review text only, or both item and review text?
    \item \textbf{RQ4:} How sensitive is RAGR to the hyperparameter settings of DPO-based task alignment?
\end{itemize}

\subsection{Experimental Settings}
\subsubsection{\textbf{Datasets}}
We evaluate \textbf{RAGR} on three Amazon review datasets~\cite{he2016ups}: \textbf{Beauty}, \textbf{Sport}, and \textbf{Toys},\footnote{\url{https://cseweb.ucsd.edu/~jmcauley/datasets/amazon/links.html}} which contain both user--item interactions and reviews. Following common practice, we filter out users and items with fewer than five interactions and chronologically order each user's interactions with their associated reviews. We adopt the leave-one-out protocol~\cite{tiger,kang2018self}, using the last interaction for testing, the second-to-last for validation, and the rest for training. Dataset statistics are reported in Table~\ref{tab:dataset}.

\begin{table}[t!]
\centering
\caption{Statistics of the datasets after preprocessing.}
\label{tab:dataset}

\resizebox{\linewidth}{!}{
    \setlength{\tabcolsep}{3.5pt}
    \renewcommand{\arraystretch}{1.05}
    \begin{tabular}{lcccccc}
    \toprule
    \textbf{Datasets} & \textbf{\#Users} & \textbf{\#Items} & \textbf{\#Inter.} & \textbf{\#RAGR Train} & \textbf{\#Val.} & \textbf{\#Test} \\
    \midrule
    Beauty & 22,363 & 12,101 & 198,502 & 285,189 & 22,363 & 22,363 \\
    Toys   & 19,412 & 11,924 & 167,597 & 238,134 & 19,412 & 19,412 \\
    Sport  & 35,598 & 18,357 & 296,337 & 414,684 & 35,598 & 35,598 \\
    \bottomrule
    \end{tabular}
}
\end{table}

\subsubsection{\textbf{Baselines}}
To comprehensively evaluate the proposed method, we compare RAGR with two groups of baselines, including representative SR and recent GR models.

\textbf{Sequential recommenders.}
\begin{itemize}[leftmargin=*]
    \item \textbf{GRU4Rec}~\cite{hidasi2015session}: a classical RNN-based SR that models user interaction sequences with gated recurrent units.
    \item \textbf{BERT4Rec}~\cite{sun2019bert4rec}: a bidirectional Transformer-based SR that learns contextual representations over user interaction sequences.
    \item \textbf{SASRec}~\cite{kang2018self}: a self-attention-based SR that captures both short-term and long-range dependencies in user histories.
    \item \textbf{S$^3$-Rec}~\cite{zhou2020s3}: a self-supervised SR that enhances sequence representation learning with auxiliary pretraining tasks.
\end{itemize}

\textbf{Generative recommenders.}
\begin{itemize}[leftmargin=*]
    \item \textbf{TIGER}~\cite{tiger}: a representative generative recommender that tokenizes items into multi-level semantic IDs and performs autoregressive next-item generation.
    \item \textbf{LETTER}~\cite{letter}: a generative recommender that improves next-item generation through learnable item tokenization.
\end{itemize}

\begin{table*}[t]
\caption{Overall comparison. \textit{Imp.} denotes the relative improvement of RAGR over its corresponding backbone model.``\textbf{$^*$}'' indicates statistically significant improvements (two-sided t-test with $p<0.05$) over the corresponding backbone model.}
\label{tab:main_results_compact}
\centering
\setlength{\tabcolsep}{5.0pt}
\renewcommand{\arraystretch}{0.75}
\resizebox{\textwidth}{!}{
\begin{tabular}{llcccc|cC{2.0cm}c|cC{2.5cm}c}
\toprule
\textbf{Dataset} & \textbf{Metric} & \textbf{GRU4Rec} & \textbf{BERT4Rec} & \textbf{SASRec} & \textbf{S$^3$-Rec} & \textbf{TIGER} & \textbf{TIGER+RAGR} & \textbf{Imp.} & \textbf{LETTER} & \textbf{LETTER+RAGR} & \textbf{Imp.} \\
\midrule

\multirow{6}{*}{Beauty}
& HIT@5   & 0.0174 & 0.0202 & 0.0377 & 0.0361 & 0.0386 & \textbf{0.0435}$^*$ & \textit{\textbf{13\%}} & 0.0371 & \textbf{0.0446}$^*$ & \textit{\textbf{20\%}} \\
& NDCG@5  & 0.0108 & 0.0111 & 0.0220 & 0.0228 & 0.0254 & \textbf{0.0292}$^*$ & \textit{\textbf{15\%}} & 0.0253 & \textbf{0.0294}$^*$ & \textit{\textbf{16\%}} \\
& HIT@10  & 0.0323 & 0.0311 & 0.0605 & 0.0598 & 0.0607 & \textbf{0.0649}$^*$ & \textit{\textbf{7\% }} & 0.0582 & \textbf{0.0677}$^*$ & \textit{\textbf{16\%}} \\
& NDCG@10 & 0.0156 & 0.0155 & 0.0303 & 0.0305 & 0.0325 & \textbf{0.0361}$^*$ & \textit{\textbf{11\%}} & 0.0321 & \textbf{0.0370}$^*$ & \textit{\textbf{15\%}} \\
& HIT@20  & 0.0513 & 0.0518 & 0.0933 & 0.0940 & 0.0866 & \textbf{0.0944}$^*$ & \textit{\textbf{9\% }} & 0.0885 & \textbf{0.1019}$^*$ & \textit{\textbf{15\%}} \\
& NDCG@20 & 0.0204 & 0.0207 & 0.0385 & 0.0391 & 0.0391 & \textbf{0.0435}$^*$ & \textit{\textbf{11\%}} & 0.0397 & \textbf{0.0455}$^*$ & \textit{\textbf{15\%}} \\
\midrule

\multirow{6}{*}{Toys}
& HIT@5   & 0.0147 & 0.0140 & 0.0369 & 0.0376 & 0.0331 & \textbf{0.0410}$^*$ & \textit{\textbf{24\%}} & 0.0321 & \textbf{0.0386}$^*$ & \textit{\textbf{20\%}} \\
& NDCG@5  & 0.0095 & 0.0082 & 0.0217 & 0.0238 & 0.0206 & \textbf{0.0259}$^*$ & \textit{\textbf{26\%}} & 0.0210 & \textbf{0.0243}$^*$ & \textit{\textbf{16\%}} \\
& HIT@10  & 0.0257 & 0.0177 & 0.0591 & 0.0604 & 0.0526 & \textbf{0.0630}$^*$ & \textit{\textbf{20\%}} & 0.0512 & \textbf{0.0604}$^*$ & \textit{\textbf{18\%}} \\
& NDCG@10 & 0.0130 & 0.0090 & 0.0289 & 0.0311 & 0.0269 & \textbf{0.0329}$^*$ & \textit{\textbf{22\%}} & 0.0272 & \textbf{0.0312}$^*$ & \textit{\textbf{15\%}} \\
& HIT@20  & 0.0420 & 0.0333 & 0.0860 & 0.0910 & 0.0781 & \textbf{0.0934}$^*$ & \textit{\textbf{20\%}} & 0.0770 & \textbf{0.0902}$^*$ & \textit{\textbf{17\%}} \\
& NDCG@20 & 0.0172 & 0.0104 & 0.0356 & 0.0388 & 0.0333 & \textbf{0.0406}$^*$ & \textit{\textbf{22\%}} & 0.0337 & \textbf{0.0387}$^*$ & \textit{\textbf{15\%}} \\
\midrule

\multirow{6}{*}{Sport}
& HIT@5   & 0.0152 & 0.0114 & 0.0212 & 0.0204 & 0.0231 & \textbf{0.0267}$^*$ & \textit{\textbf{16\%}} & 0.0240 & \textbf{0.0264}$^*$ & \textit{\textbf{10\%}} \\
& NDCG@5  & 0.0102 & 0.0085 & 0.0138 & 0.0136 & 0.0148 & \textbf{0.0176}$^*$ & \textit{\textbf{19\%}} & 0.0156 & \textbf{0.0172}$^*$ & \textit{\textbf{10\%}} \\
& HIT@10  & 0.0231 & 0.0224 & 0.0311 & 0.0306 & 0.0385 & \textbf{0.0415}$^*$ & \textit{\textbf{8\% }} & 0.0403 & \textbf{0.0433}$^*$ & \textit{\textbf{ 7\%}} \\
& NDCG@10 & 0.0127 & 0.0091 & 0.0177 & 0.0169 & 0.0197 & \textbf{0.0224}$^*$ & \textit{\textbf{14\%}} & 0.0209 & \textbf{0.0227}$^*$ & \textit{\textbf{ 9\%}} \\
& HIT@20  & 0.0359 & 0.0300 & 0.0506 & 0.0483 & 0.0564 & \textbf{0.0619}$^*$ & \textit{\textbf{10\%}} & 0.0612 & \textbf{0.0661}$^*$ & \textit{\textbf{ 8\%}} \\
& NDCG@20 & 0.0159 & 0.0080 & 0.0220 & 0.0213 & 0.0242 & \textbf{0.0274}$^*$ & \textit{\textbf{13\%}} & 0.0262 & \textbf{0.0284}$^*$ & \textit{\textbf{ 9\%}} \\
\bottomrule
\end{tabular}
}
\end{table*}

\subsubsection{\textbf{Evaluation Metrics}}

Following common practice in GR~\cite{tiger,kang2018self}, we evaluate all methods with top-$K$ ranking metrics, including \textbf{HIT@K} and \textbf{NDCG@K}, where $K \in \{5,10,20\}$. 

\subsubsection{\textbf{Implementation Details}}
To verify the effectiveness and generality of RAGR, we instantiate it on two representative models, \textbf{TIGER+RAGR} and \textbf{LETTER+RAGR}. All baselines are trained and evaluated under the same data split and protocol: conventional sequential recommenders use only item interaction sequences following their original settings, while GR adopt their standard item tokenization and autoregressive prediction. 
In the \textbf{tokenizer training} stage, we use T5\footnote{\url{https://huggingface.co/sentence-transformers/sentence-t5-base}} to encode items and reviews. The RQ-VAE uses a six-layer encoder/decoder MLP with hidden dimensions $[2048,1024,512,256,128,64]$, $4$ residual codebooks each with $256$ codewords of dimension $32$, and k-means initialization ($100$ iterations, $10$ clusters). It is optimized with AdamW (learning rate $1\times10^{-3}$, batch size $2048$, weight decay $1\times10^{-4}$) for at most $2000$ epochs with collision-rate evaluation.

For \textbf{review-augmented user sequence modeling}, we interleave item and review semantic IDs chronologically, with a maximum history length of $20$ (longer histories truncated from the left). The GR backbone uses T5~\cite{raffel2020t5}, trained for $200$ epochs with AdamW (learning rate $1\times10^{-3}$, per-device batch size $256$, gradient accumulation $2$, weight decay $0.01$) under a cosine scheduler with $0.01$ warmup ratio. 
For \textbf{item-centric task generation alignment}, we initialize the policy model from the trained GR model and fix the reference model during DPO; the $\beta$ in Eq.~(\ref{eq:dpo}) is set to $0.5$--$0.7$ (sensitivity analysis in Section~\ref{sec:dpo_para}), optimized with AdamW (learning rate $1\times10^{-6}$, per-device batch size $256$). All experiments run on an Ubuntu server with eight NVIDIA RTX PRO 6000 GPUs (96\,GB); main results are averaged over three runs with seeds $42$, $43$, and $44$.
We adopt two-stage training: $L_{\mathrm{seq}}$ for next-item and next-review prediction, followed by $L_{\mathrm{DPO}}$ for item-centric alignment.
The algorithm is provided in Appendix~\ref{app:algo} and the sid sensitivity experiment is provided in Appendix~\ref{app:sid_sen}.

\subsection{Overall Performance (RQ1)}

To answer RQ1, we compare RAGR against SR and GR baselines; Table~\ref{tab:main_results_compact} reports the overall results, from which several observations can be drawn.
\textbf{First, RAGR consistently improves both GR backbones across all datasets and metrics.} Specifically, it achieves relative gains of \textit{7\%--26\%} for \textit{TIGER} and \textit{7\%--20\%} for \textit{LETTER}, demonstrating its effectiveness and generalizability across different GR architectures.
\textbf{Second, RAGR also achieves the best overall performance against conventional sequential recommenders.} On all three datasets, both TIGER+RAGR and LETTER+RAGR outperform GRU4Rec, BERT4Rec, SASRec, and S$^3$-Rec on every metric---notable given that these baselines span strong recurrent, Transformer-based, and self-supervised paradigms. This suggests that incorporating review feedback into the generative sequence provides predictive signals beyond item-only interaction modeling.
\textbf{Third, review-augmented modeling is especially beneficial when textual feedback carries stronger preference signals.} This is most evident on Toys, where RAGR's gains consistently exceed those on Beauty and Sport. 

\begin{figure*}[ht]
    \centering
    \includegraphics[width=0.95\textwidth]{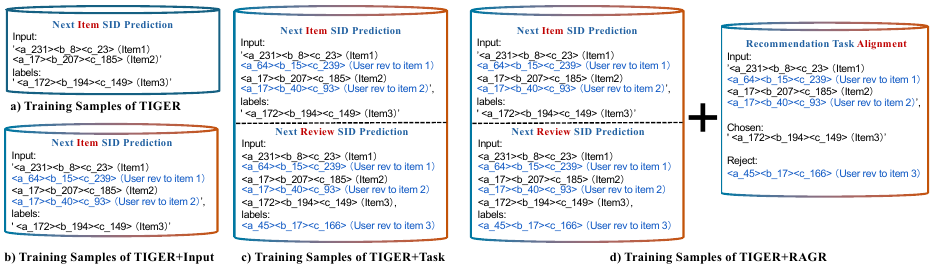}
    \caption{Illustration of the progressively enhanced training paradigms in the ablation study.}
    \label{fig:ablation}
\end{figure*}

\subsection{Ablation Study (RQ2)}
To answer \textbf{RQ2} and understand the contribution of each component in RAGR, we conduct an ablation study. As illustrated in Fig.~\ref{fig:ablation}, we progressively extend the training paradigm from the original item-only generative recommendation to the full RAGR framework. Specifically, the compared variants are defined as follows:
\begin{itemize}[leftmargin=*]
    \item \textbf{+Input}: The original item-only input sequence is augmented with review semantic IDs, while the original next-item prediction target remains unchanged.
    \item \textbf{+Task}: Review prediction is further introduced into the textual generation process, so that both next-item SID prediction and next-review SID prediction are jointly modeled within the same unified sequence generation framework.
    \item \textbf{+RAGR}: The proposed item-centric task alignment is further incorporated, where the target item SID is treated as the chosen output and the corresponding review SID is treated as the rejected output in DPO training.
\end{itemize}

\begin{table*}[t]
\caption{Ablation study of input expansion, task expansion, and task alignment on Beauty, Toys, and Sport.}
\label{tab:ablation_sft_dpo}
\centering
\setlength{\tabcolsep}{5pt}
\renewcommand{\arraystretch}{0.85}
\begin{threeparttable}
\resizebox{\textwidth}{!}{
\begin{tabular}{lcccccc|cccccc|cccccc}
\toprule
\multirow{2}{*}{\textbf{Method}} 
& \multicolumn{6}{c|}{\textbf{Beauty}} 
& \multicolumn{6}{c|}{\textbf{Toys}} 
& \multicolumn{6}{c}{\textbf{Sport}} \\
\cmidrule(lr){2-7} \cmidrule(lr){8-13} \cmidrule(lr){14-19}
& H@5 & N@5 & H@10 & N@10 & H@20 & N@20
& H@5 & N@5 & H@10 & N@10 & H@20 & N@20
& H@5 & N@5 & H@10 & N@10 & H@20 & N@20 \\
\midrule

TIGER
& 0.0386 & 0.0254 & 0.0607 & 0.0325 & 0.0866 & 0.0391
& 0.0331 & 0.0206 & 0.0526 & 0.0269 & 0.0781 & 0.0333
& 0.0231 & 0.0148 & 0.0385 & 0.0197 & 0.0564 & 0.0242 \\

+Input
& 0.0260 & 0.0164 & 0.0417 & 0.0215 & 0.0619 & 0.0265
& 0.0273 & 0.0174 & 0.0424 & 0.0223 & 0.0676 & 0.0287
& 0.0191 & 0.0120 & 0.0317 & 0.0161 & 0.0482 & 0.0202 \\

+Task
& 0.0427 & 0.0289 & 0.0649 & 0.0361 & 0.0938 & 0.0434
& 0.0405 & 0.0258 & 0.0624 & 0.0327 & 0.0930 & 0.0402
& 0.0266 & 0.0170 & 0.0413 & 0.0223 & 0.0606 & 0.0271 \\

\textbf{+RAGR}
& \textbf{0.0435} & \textbf{0.0292} & \textbf{0.0649} & \textbf{0.0361} & \textbf{0.0944} & \textbf{0.0435}
& \textbf{0.0410} & \textbf{0.0259} & \textbf{0.0630} & \textbf{0.0329} & \textbf{0.0934} & \textbf{0.0406}
& \textbf{0.0267} & \textbf{0.0176} & \textbf{0.0415} & \textbf{0.0224} & \textbf{0.0619} & \textbf{0.0274} \\

\midrule

LETTER
& 0.0371 & 0.0253 & 0.0582 & 0.0321 & 0.0885 & 0.0397
& 0.0321 & 0.0210 & 0.0512 & 0.0272 & 0.0770 & 0.0337
& 0.0240 & 0.0156 & 0.0403 & 0.0209 & 0.0612 & 0.0262 \\

+Input
& 0.0311 & 0.0205 & 0.0511 & 0.0269 & 0.0778 & 0.0336
& 0.0273 & 0.0174 & 0.0424 & 0.0223 & 0.0676 & 0.0287
& 0.0194 & 0.0126 & 0.0310 & 0.0164 & 0.0484 & 0.0208 \\

+Task
& 0.0421 & 0.0280 & 0.0634 & 0.0353 & 0.0971 & 0.0435
& 0.0382 & 0.0240 & 0.0600 & 0.0310 & 0.0901 & 0.0386
& 0.0263 & 0.0172 & 0.0430 & 0.0226 & 0.0658 & 0.0284 \\

\textbf{+RAGR}
& \textbf{0.0446} & \textbf{0.0294} & \textbf{0.0677} & \textbf{0.0370} & \textbf{0.1019} & \textbf{0.0455}
& \textbf{0.0386} & \textbf{0.0243} & \textbf{0.0604} & \textbf{0.0312} & \textbf{0.0902} & \textbf{0.0387}
& \textbf{0.0264} & \textbf{0.0172} & \textbf{0.0433} & \textbf{0.0227} & \textbf{0.0661} & \textbf{0.0284} \\

\bottomrule
\end{tabular}
}
\end{threeparttable}
\end{table*}

\begin{figure*}[h!]
\centering
\includegraphics[width=0.95\textwidth]{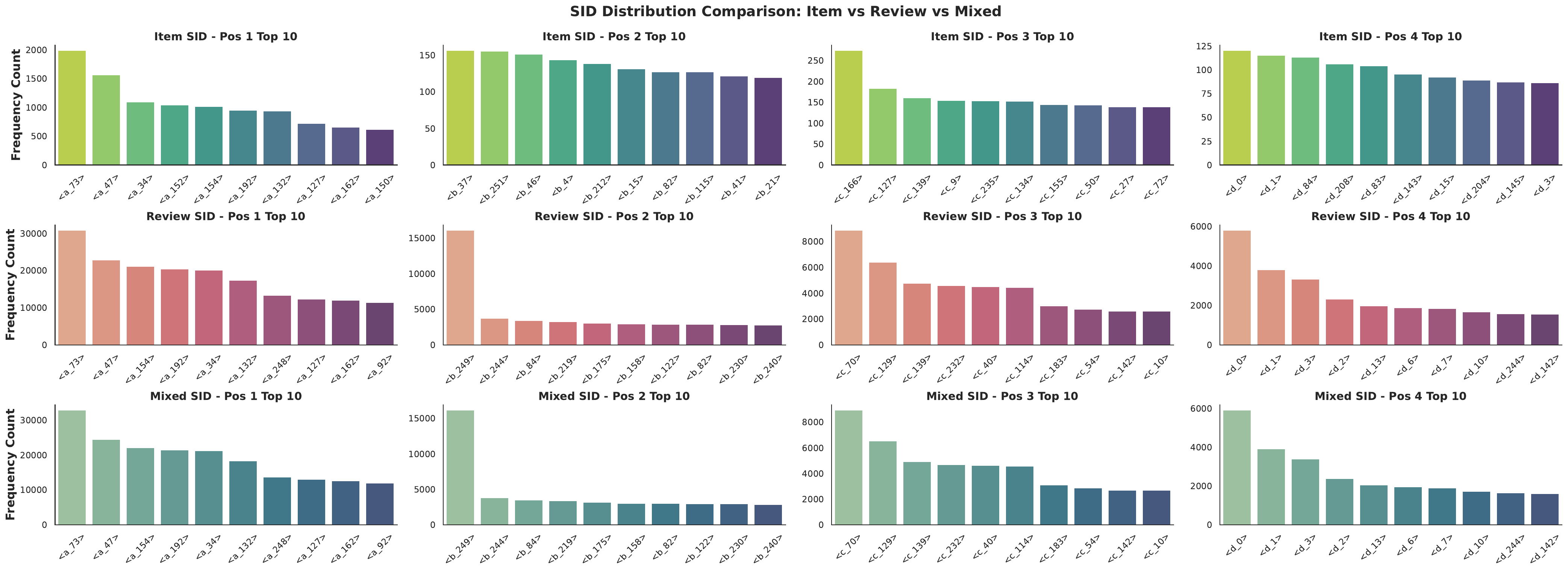}
\caption{Comparison of Top-10 SID frequency distributions across item-only, review-only, and mixed item-review sequences.}
\label{fig:sid_dist}
\end{figure*}

The results are reported in Table~\ref{tab:ablation_sft_dpo}, from which several observations can be made.
\textbf{First, simply expanding the input sequence with review signals (+Input) does not improve performance; instead, it consistently degrades both TIGER and LETTER on all three datasets.} On Beauty, TIGER drops from 0.0386 to 0.0260 in HIT@5 and LETTER from 0.0371 to 0.0311, with the same trend on Toys and Sport. Inserting review SIDs while keeping the next-item objective unchanged leaves the model untrained to interpret or predict them, so the added tokens disturb item-transition patterns and act more like noise. Review augmentation thus requires a matched training objective, not a trivial input-level extension.
\textbf{Second, extending the training objective to include review prediction (+Task) improves performance substantially over both the backbone and +Input.} On Toys, TIGER+{Task} reaches 0.0405/0.0258 on HIT@5/NDCG@5, clearly outperforming TIGER and TIGER+{Input}, with LETTER+{Task} and the results on Beauty and Sport showing the same gains. By requiring the model to \emph{generate} review SIDs within unified sequence modeling, +Task turns them into meaningful targets and lets the model learn the correspondence between interactions and feedback. Review SIDs thus help only when explicitly incorporated into the generative task.
\textbf{Third, adding item-centric task alignment further improves performance in most settings and yields the best overall results.} On Beauty, TIGER+{RAGR} improves over TIGER+{Task} from 0.0427 to 0.0435 in HIT@5, and LETTER+{RAGR} from 0.0421 to 0.0446 in HIT@5 and 0.0971 to 0.1019 in HIT@20; gains are consistent on Toys and stable, if smaller, on Sport. Task expansion alone cannot resolve the tension of mixed generation, as the model may divert capacity to review SIDs. By applying DPO-based alignment that prefers item over review SIDs at prediction positions, RAGR keeps reviews as enriching context while anchoring optimization to item recommendation---confirming that alignment turns review prediction into an auxiliary rather than competing objective.

To understand \emph{why} review augmentation helps, we compare the Top-10 SID frequency distributions of item-only, review-only, and mixed sequences across four token positions (Figure~\ref{fig:sid_dist}). Item and review SIDs differ markedly: item distributions concentrate on a few recurring SIDs (notably at the first and third positions), while review SIDs concentrate on a different high-frequency set, showing that reviews encode user-side evaluative semantics rather than redundant item descriptions. The mixed sequence inherits both---activating additional review-related SIDs and reshaping the frequency profile---so review augmentation changes the sequence's semantic composition rather than merely lengthening it, exposing the model to both behavioral outcomes (item SIDs) and preference rationales (review SIDs). This explains the gains of \textbf{+Task} and \textbf{+RAGR}: turning the item-only sequence into a mixed behavioral-semantic one lets the autoregressive model learn from post-interaction feedback, so the improvements stem not only from extra textual information but also from better use of the unified SID space.

\begin{figure*}[ht]
\centering
\includegraphics[width=0.95\textwidth]{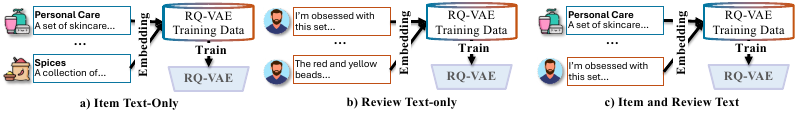}
\caption{Illustration of the three tokenizer: training RQ-VAE on item text only, review text only, and both item and review text.}
\label{fig:tok_exp}
\end{figure*}

\begin{figure*}[ht]
\centering
    \includegraphics[width=0.95\textwidth]{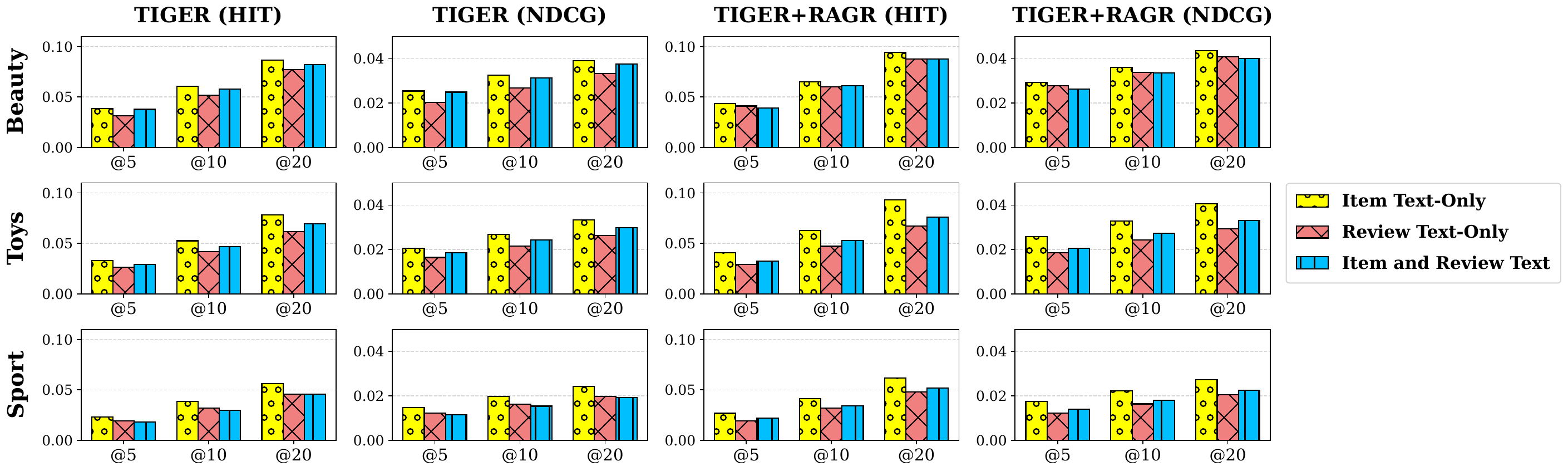}
    \caption{Impact of different tokenizer training strategies on downstream recommendation performance. We compare tokenizers trained on \emph{Item Text-Only}, \emph{Review Text-Only}, and \emph{Item and Review Text} under both TIGER and TIGER+RAGR.}
    \label{fig:tok_exp_result}
\end{figure*}

\subsection{Tokenizer Experiment (RQ3)}

To answer RQ3, we consider three tokenizer training strategies, as illustrated in Fig.~\ref{fig:tok_exp}. Since the proposed framework relies on a unified semantic ID space for both item interactions and review feedback, the quality of the tokenizer is critical to the effectiveness of subsequent review-augmented sequence modeling and item-centric task alignment.
\begin{itemize}[leftmargin=*]
    \item \textbf{Item Text-Only}: the RQ-VAE tokenizer is trained only on item text embeddings.
    \item \textbf{Review Text-Only}: the RQ-VAE tokenizer is trained only on review text embeddings.
    \item \textbf{Item and Review Text}: the RQ-VAE tokenizer is trained on the combination of item text and review text embeddings.
\end{itemize}

\begin{table}[ht]
\centering
\caption{Comparison of different tokenizer training strategies in terms of training efficiency and collision rate (Col.).}
\label{tab:tok_exp}
\setlength{\tabcolsep}{2pt}
\renewcommand{\arraystretch}{1.0}
\resizebox{\linewidth}{!}{
\begin{tabular}{lcccc}
\toprule
\textbf{Tokenizer} & \textbf{Training Time} & \textbf{Beauty Col.} & \textbf{Toys Col.} & \textbf{Sport Col.} \\
\midrule
Item Text-Only       & 2s/epoch  & 0.0009 & 0.0011 & 0.0025 \\
Review Text-Only     & 29s/epoch & 0.0009 & 0.0012 & 0.0024 \\
Item and Review Text & 33s/epoch & 0.0008 & 0.0044 & 0.0005 \\
\bottomrule
\end{tabular}
}
\end{table}

To provide a more direct comparison, Table~\ref{tab:tok_exp} summarizes the training efficiency and collision rates of the three tokenizer training strategies across the three datasets. 
Overall, the tokenizer experiment leads to three main observations. 

First, \textbf{Item Text-Only} consistently achieves the strongest downstream recommendation performance in most settings, for both the original TIGER and the review-augmented TIGER+RAGR variant. As shown in Fig.~\ref{fig:tok_exp_result}, the item-text tokenizer yields the highest HIT and NDCG values on Beauty, Toys, and Sport in the majority of cases. This suggests that, for generative recommendation, learning semantic IDs from item text provides a more stable and effective token space for next-item prediction.
Second, \textbf{Review Text-Only} generally performs the worst among the three strategies. Although the review text contains rich user preference signals, using it alone to train the tokenizer appears to shift the item SID distribution, which weakens the quality of item representations and harms downstream recommendation performance. In contrast, \textbf{Item and Review Text} often performs better than \textbf{Review Text-Only}, but still does not consistently surpass \textbf{Item Text-Only}. This indicates that directly mixing review text into tokenizer training does not necessarily lead to a better semantic ID space, even though review feedback is beneficial at the sequence modeling stage.
Third, Table~\ref{tab:tok_exp} shows that \textbf{Item Text-Only} is also substantially more efficient, requiring only $2$s per epoch, compared with $29$s and $33$s for \textbf{Review Text-Only} and \textbf{Item and Review Text}, respectively. Although the collision rates do not always perfectly align with downstream recommendation performance, the item-text tokenizer provides the best overall trade-off between efficiency and effectiveness. Therefore, in the remaining experiments, we adopt \textbf{Item Text-Only} as the default tokenizer training strategy.

\subsection{DPO Sensitivity Experiment (RQ4)}

\label{sec:dpo_para}
\begin{figure}[h!]
\centering
\includegraphics[width=0.45\textwidth]{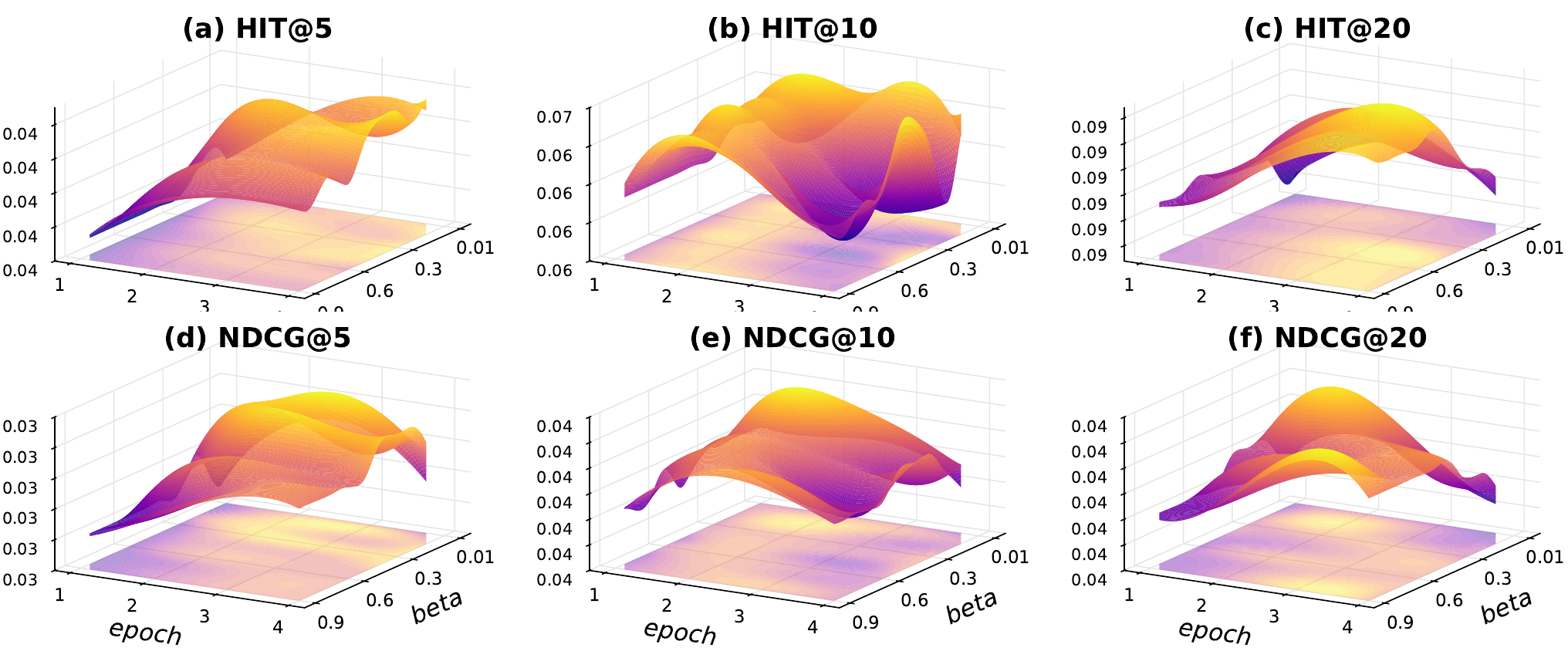}
\caption{Sensitivity analysis of the DPO-based alignment with respect to the coefficient $\beta$ and training epoch.}
\label{fig:dpo_sensitivity}
\end{figure}
To answer \textbf{RQ4}, we study how sensitive RAGR is to the DPO alignment hyperparameters, focusing on the preference coefficient $\beta$, which controls the sharpness of preference optimization, and the training epoch, which determines how far the GR backbone is aligned toward item-centric generation. Since DPO is introduced to preserve the \emph{next-item recommendation} boundary after review signals enter the sequence, its sensitivity directly reflects the stability of the alignment. Figure~\ref{fig:dpo_sensitivity} visualizes the performance landscape over different $\beta$--epoch combinations on HIT@$\{5,10,20\}$ and NDCG@$\{5,10,20\}$.
The surface is generally smooth, indicating stable alignment over a broad range of settings. For most metrics, moderate $\beta$ outperforms overly small or large values: too small a $\beta$ leaves the preference signal too weak to enforce item-centric alignment, while too large a $\beta$ overemphasizes the chosen--rejected contrast and harms generative capability. Similarly, the best results usually appear within a moderate training range rather than at the earliest or latest epochs. Overall, RAGR does not rely on an extremely narrow hyperparameter region yet still benefits from proper tuning; we accordingly select the $\beta$--epoch combination with the best performance--stability trade-off on the validation set as the default configuration.

\section{Related work}
\label{sec:relatedwork}
\subsection{Sequential Recommendation}
Sequential recommendation (SR) models users' evolving preferences from historical interaction sequences to predict the next item, and is a fundamental paradigm in recommender systems~\cite{pan2026survey,wang2023sequential} studied mainly under the \emph{next-item prediction} formulation. Methods differ in how they encode the sequence: early recurrent architectures such as GRU4Rec~\cite{hidasi2015session} model transitions via hidden states, while Transformer-based models---SASRec~\cite{kang2018self} and BERT4Rec~\cite{sun2019bert4rec}---use self-attention and bidirectional context to capture long-range dependencies. Self-supervised and contrastive objectives further improve robustness, as in S$^3$-Rec~\cite{zhou2020s3} and contrastive approaches~\cite{xie2022contrastive}, and richer inductive biases have been explored, including temporal modeling~\cite{ye2020time}, dynamic graphs~\cite{zhang2022dynamic}, hyper adapters~\cite{li2023hamur,lan2023m3rec}, and LLM-enhanced representations~\cite{liu2025llmemb,wang2025rethinking}. \textbf{Despite these advances, most SR methods model behavior as an item-only sequence, capturing what users selected while overlooking richer signals such as reviews that explain why items were selected or rejected.} In contrast, our work incorporates review feedback into the generative user sequence for richer preference modeling.

\subsection{Generative Recommendation}
Generative recommendation (GR) reformulates recommendation as producing the target item through autoregressive decoding rather than ranking over a fixed vocabulary, a rapidly growing direction whose progress recent surveys attribute to tokenization, generative architecture, and optimization~\cite{deldjoo2024review,li2025survey,hou2025towards, wang2026gflowgr,jia2026glorank,hu2026think}. A representative line learns \emph{semantic IDs} for items and generates the next item over these discrete codes, such as TIGER~\cite{tiger} and LETTER~\cite{letter}, avoiding classification over an extremely large item vocabulary. Subsequent studies advance tokenization and decoding, including ID tokenization for LLM-based recommendation~\cite{tokenrec,ju2025generative}, parallel decoding of long semantic IDs~\cite{hou2025generating}, contextual tokenization of action sequences~\cite{hou2025actionpiece}, unified sparse-dense generation~\cite{yang2025sparse}, acceleration-oriented decoding~\cite{wang2025nezha}, and combining semantic and collaborative codes~\cite{UNGER}. \textbf{Nonetheless, existing GR methods still model the user sequence as an item-only trajectory with the next item as the generation target.} By replacing raw IDs with semantic IDs, GR mainly changes \emph{how} items are represented rather than \emph{what} constitutes the user sequence---enhancing item representation while leaving the user sequence largely untouched, a gap our work addresses by incorporating review feedback.

\subsection{Review-Aware Recommendation}
Review-aware recommendation leverages user-generated reviews to enrich preference modeling~\cite{wen2026memory,zhang2026evoking} beyond interaction signals, as reviews carry rich information about user opinions, product attributes, and decision rationale~\cite{wu2022survey,Hasan2025review}. A large body of work uses reviews as auxiliary text for matching or rating prediction: early joint user--item modeling~\cite{zheng2017joint}, aspect-aware modeling~\cite{cheng2018aspect,nie2026hadsf}, review-level explanation~\cite{chen2018neural,dong2020asymmetrical}, review property modeling~\cite{wang2021leveraging}, graph- and hypergraph-based interaction~\cite{shi2018heterogeneous,AHAG,wang2020traffic}, and critical analyses of reviews' utility~\cite{sachdeva2020useful}. More recent LLM-based approaches exploit text understanding~\cite{wu2025empowering, wu2026deepresearch,dong2026learning, dong2026utilize, dong2025threshold, zhang2026tearag} and reasoning, such as recommendation-as-language-processing~\cite{geng2022recommendation}, prompt-based personalization~\cite{lyu2024llm,jia2024mill,jia2024mill,li2026survey}, rationale enhanced recommendation~\cite{wang2024rdrec}, and review-analysis-based feature enhancement~\cite{assi2024llm}. \textbf{However, most methods treat reviews as side information used to enhance representations or matching, rather than integrating them into a unified generative sequence with item interactions.} In contrast, our work maps reviews into the same SID space as items, integrates them into the generative user sequence, and adds item-centric task alignment so review signals enhance next-item generation without becoming competing targets.
\section{Conclusion}
\label{sec:conclusion}
In this paper, we rethink GR beyond the item-only paradigm, arguing that modeling user sequences solely as item trajectories captures \emph{what} users choose but overlooks \emph{why}. To our knowledge, RAGR is the first work to bring reviews into GR by encoding them as SID tokens within the user sequence.
Experiments on three real-world datasets show that RAGR consistently outperforms strong item-only baselines across multiple backbones, with ablations confirming both components are indispensable. In the future, we hope this work inspires broader exploration of richer user feedback---such as search queries, comments, and multimodal interactions---within a unified generative framework for more comprehensive user modeling.

\section*{Ethical Considerations}
By turning user reviews into the generation signal, RAGR inherits both the promise and the responsibility of learning from what people write. Although we rely solely on public Amazon benchmarks with no author-collected personal data, three concerns merit attention. First, reviews carry traces of individual voice and opinion; RAGR mitigates this by consuming reviews only as quantized semantic IDs and never storing or emitting raw review text at inference, and we recommend anonymization and consent safeguards in real deployments. Second, reviews are unevenly distributed, so augmentation may amplify popularity and exposure bias toward well-reviewed items at the expense of long-tail content—a risk that calls for debiasing objectives and regular auditing across item and user groups. Third, because generation is review-conditioned, the model is in principle exposed to fake or adversarial reviews, which can be countered with spam detection and robust training. Used as intended, RAGR's remaining risks are those common to all recommenders, such as filter bubbles, and are best managed through diversity-aware deployment and human oversight.

\begin{acks}
This research was supported by the National Natural Science Foundation of China (NSFC) under Grant Nos. 72672019, 72231010, and 62502404, and was supported in part by the Hong Kong Research Grants Council (Research Impact Fund No.R1015-23, Collaborative Research Fund No.C1043-24GF, RGC Research Fellow Scheme No.RFS2627-1S03, General Research Fund No.11218325, No.11212926), Institute of Digital Medicine of City University of Hong Kong (No.9229503), and Huawei (Huawei Innovation Research Program).
\end{acks}

\balance
\bibliographystyle{ACM-Reference-Format}
\bibliography{chapter/6ref}

\appendix
\section{Overall Training Procedure of RAGR}
\label{app:algo}
\begin{algorithm}[t]
\caption{Overall Training Procedure of RAGR}
\label{alg:ragr}
\small
\begin{algorithmic}[1]
\Require Item text corpus $\mathcal{X}^{I}$, review text corpus $\mathcal{X}^{R}$, user interactions $\mathcal{S}$, and GR backbone $\pi_{\theta}$
\Ensure Trained review-augmented generative recommender $\pi_{\theta}^{*}$

\Statex \textbf{Stage 1: Tokenizer Training}
\State Encode $\mathcal{X}^{I}$ and $\mathcal{X}^{R}$ with a shared text encoder
\State Train RQ-VAE on item embeddings to obtain tokenizer $\mathcal{T}$
\State Quantize item and review embeddings with $\mathcal{T}$ to obtain $\mathbf{z}(i)$ and $\mathbf{z}(r)$
\State Freeze $\mathcal{T}$ for subsequent stages

\Statex \textbf{Stage 2: Review-Augmented User Sequence Modeling}
\State $\mathcal{D}_{\mathrm{seq}} \gets \emptyset$
\For{each user sequence $\mathcal{S}_u=[(i_t,r_t)]_{t=1}^{T_u}$}
    \State $\mathbf{h} \gets [\mathbf{z}(i_1),\mathbf{z}(r_1)]$
    \For{$t=2$ to $T_u$}
        \State Append $(\mathbf{h},\mathbf{z}(i_t))$ to $\mathcal{D}_{\mathrm{seq}}$
        \State $\mathbf{h}^{I} \gets [\mathbf{h},\mathbf{z}(i_t)]$
        \State Append $(\mathbf{h}^{I},\mathbf{z}(r_t))$ to $\mathcal{D}_{\mathrm{seq}}$
        \State $\mathbf{h} \gets [\mathbf{h}^{I},\mathbf{z}(r_t)]$
    \EndFor
\EndFor
\State Train $\pi_{\theta}$ on $\mathcal{D}_{\mathrm{seq}}$ using the autoregressive generation objective

\Statex \textbf{Stage 3: Item-Centric Task Generation Alignment}
\State $\mathcal{D}_{\mathrm{dpo}} \gets \emptyset$
\For{each user sequence $\mathcal{S}_u=[(i_t,r_t)]_{t=1}^{T_u}$}
    \State $\mathbf{h} \gets [\mathbf{z}(i_1),\mathbf{z}(r_1)]$
    \For{$t=2$ to $T_u$}
        \State $\mathbf{y}^{+} \gets \mathbf{z}(i_t)$
        \State $\mathbf{y}^{-} \gets \mathbf{z}(r_t)$
        \State Append $(\mathbf{h},\mathbf{y}^{+},\mathbf{y}^{-})$ to $\mathcal{D}_{\mathrm{dpo}}$
        \State $\mathbf{h} \gets [\mathbf{h},\mathbf{z}(i_t),\mathbf{z}(r_t)]$
    \EndFor
\EndFor
\State Initialize $\pi_{\mathrm{ref}}$ as a frozen copy of $\pi_{\theta}$
\State Optimize $\pi_{\theta}$ on $\mathcal{D}_{\mathrm{dpo}}$ using the DPO objective with fixed $\pi_{\mathrm{ref}}$
\State $\pi_{\theta}^{*} \gets \pi_{\theta}$
\State \Return $\pi_{\theta}^{*}$
\end{algorithmic}
\end{algorithm}

Algorithm~\ref{alg:ragr} summarizes the three-stage training procedure of RAGR. First, an RQ-VAE tokenizer trained on item text embeddings maps both items and reviews into a shared semantic ID space. Second, the GR backbone is trained on two autoregressive tasks: next-item generation conditioned on the review-augmented interaction history, and next-review generation conditioned on the same history and the target item. Finally, item-centric task generation alignment applies DPO to prefer the target item IDs over the corresponding review IDs under the same historical context, using a frozen copy of the sequence-trained model as the reference policy.

\section{SID Sensitivity Experiment}
\label{app:sid_sen}

\begin{table}[h]
\centering
\caption{Collision rates under different SID numbers on the three datasets.}
\label{tab:sid_collision}
\setlength{\tabcolsep}{6pt}
\renewcommand{\arraystretch}{1.05}
\begin{tabular}{cccc}
\toprule
\textbf{SID Num} & \textbf{Beauty Col.} & \textbf{Toys Col.} & \textbf{Sport Col.} \\
\midrule
3 & 0.0121 & 0.0017 & 0.0045 \\
4 & 0.0009 & 0.0012 & 0.0026 \\
5 & 0.0007 & 0.0003 & 0.0018 \\
\bottomrule
\end{tabular}
\end{table}

\begin{figure*}[tb]
\centering
\includegraphics[width=0.90\textwidth]{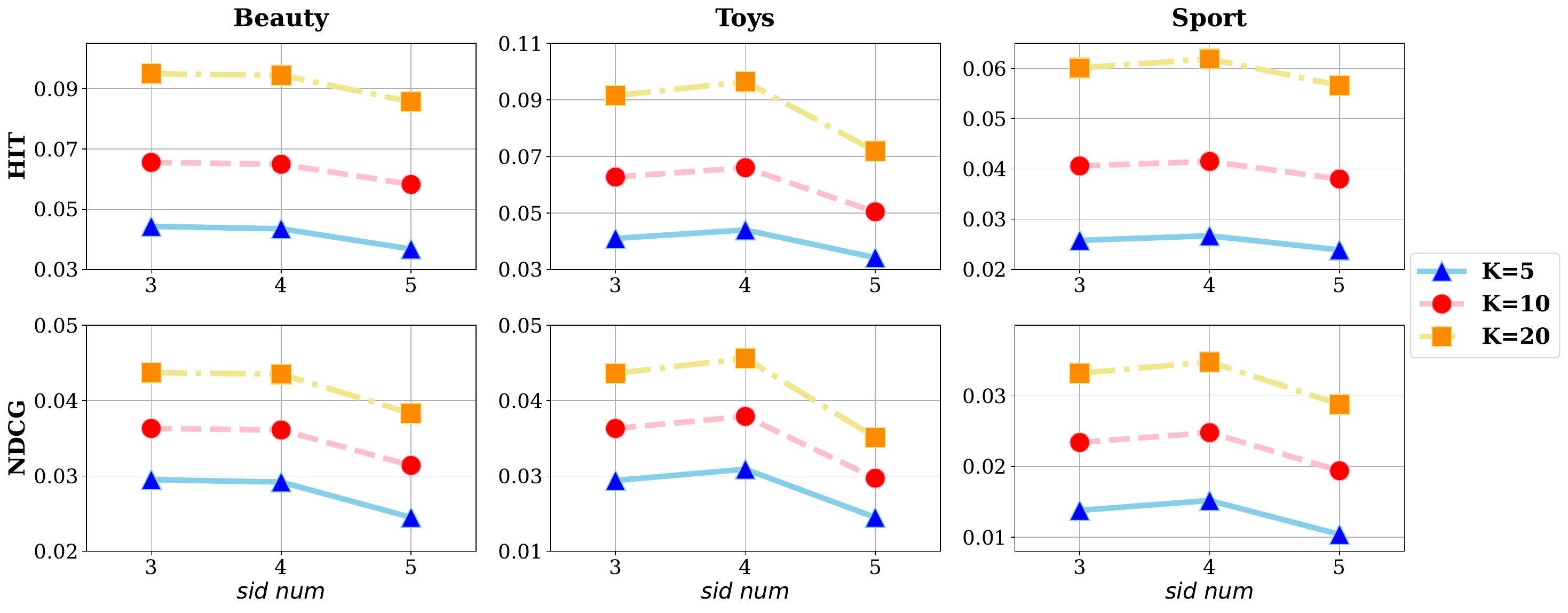}
\caption{Impact of the number of semantic ID tokens on recommendation performance. We report HIT@K and NDCG@K ($K \in \{5,10,20\}$) on Beauty, Toys, and Sport under different SID numbers.}
\label{fig:sid_sensitivity}
\end{figure*}

We investigate the impact of SID length on RAGR by varying the number of semantic ID tokens from $3$ to $5$. Intuitively, fewer tokens lead to a smaller semantic space and a higher risk of collisions, whereas more tokens improve semantic distinctiveness but increase autoregressive generation difficulty. As shown in Table~\ref{tab:sid_collision}, increasing the SID number consistently reduces collision rates across all datasets. For example, on Beauty, the collision rate decreases from $0.0121$ with $3$ tokens to $0.0009$ and $0.0007$ with $4$ and $5$ tokens, respectively. However, lower collision rates do not necessarily translate into better recommendation performance.

Figure~\ref{fig:sid_sensitivity} reveals three main observations. \emph{First}, using $4$ semantic ID tokens achieves the best overall HIT and NDCG performance across the three datasets, suggesting an effective balance between semantic distinctiveness and generation difficulty. \emph{Second}, shorter SIDs increase collisions and weaken item-level discriminability. Interestingly, using $3$ tokens still yields relatively competitive results on Beauty, possibly because shared semantic IDs introduce a smoothing effect that partially alleviates data sparsity during generation. However, this potential benefit comes at the cost of reduced item distinguishability. \emph{Third}, increasing the SID number from $4$ to $5$ further reduces collisions but consistently degrades recommendation performance, suggesting that longer SIDs enlarge the generation space and increase decoding difficulty without commensurate gains in representation quality. Overall, these results reveal a trade-off between semantic uniqueness and generation complexity. Accordingly, we adopt $4$ semantic ID tokens as the default configuration of RAGR.

\end{document}